\documentclass[twocolumn]{aastex631}

\usepackage{mathrsfs}
\usepackage{amssymb}
\usepackage{amsmath}
\usepackage{graphicx}
\usepackage{amsfonts,amsbsy}
\usepackage{bm}
\usepackage{nicefrac}
\usepackage{xcolor}
\usepackage[capitalise]{cleveref}
\usepackage{xcolor}

\makeatletter
\g@addto@macro\bfseries{\boldmath}
\makeatother

\definecolor{lcolor}{rgb}{0.5,0,0}
\definecolor{citcolor}{rgb}{0,0.3,0.0}
\definecolor{ao(english)}{rgb}{0.0, 0.5, 0.0}
\usepackage{comment}

\newcommand{\EoS}{{\rm EoS}}
\renewcommand{\tilde}{\widetilde}
\renewcommand{\epsilon}{\varepsilon}
\renewcommand{\log}{\ln}

\begin{document}

\title{\texorpdfstring{\emph{Ab-initio} QCD calculations impact the inference of \\ the neutron-star-matter equation of state}{\emph{Ab-initio} QCD calculations impact the inference of the neutron-star-matter equation of state} }

\author[0000-0003-3469-7574]{Tyler Gorda}
\affiliation{Technische Universit\"{a}t Darmstadt, Department of Physics, 64289 Darmstadt, Germany}
\affiliation{ExtreMe Matter Institute EMMI and Helmholtz Research Academy for FAIR, GSI Helmholtzzentrum f\"ur Schwerionenforschung GmbH, 64291 Darmstadt, Germany}
\author[0000-0002-2188-3549]{Oleg Komoltsev}
\affiliation{Faculty of Science and Technology, University of Stavanger, 4036 Stavanger, Norway}
\author[0000-0001-7991-3096]{Aleksi Kurkela}
\affiliation{Faculty of Science and Technology, University of Stavanger, 4036 Stavanger, Norway}
\begin{abstract}
We demonstrate that \emph{ab-initio} calculations in QCD at high densities offer significant and nontrivial information about the equation of state of matter in the cores of neutron stars, going beyond that which is obtainable from current astrophysical observations. We do so by extrapolating the equation of state to neutron-star densities using a Gaussian process and conditioning it sequentially with astrophysical observations and QCD input. Using our recent work, imposing the latter does not require an extrapolation to asymptotically high density. We find the QCD input to be complementary to the astrophysical observations, offering strong additional constraints at the highest densities reached in the cores of neutron stars; with the QCD input, the equation of state is no longer prior dominated at any density. The QCD input reduces the pressure and speed of sound at high densities, and it predicts that binary collisions of equal-mass neutron stars will produce a black hole with greater than $95\%$ ($68\%$) credence for masses $M \geq 1.38 M_\odot$ ($M \geq 1.25 M_\odot$). We provide a Python implementation  of the QCD likelihood function so that it can be conveniently used within other inference setups.
\end{abstract}

\section{Introduction}

\begin{figure}[ht!]
    \centering
    \includegraphics[width = 0.43\textwidth ]{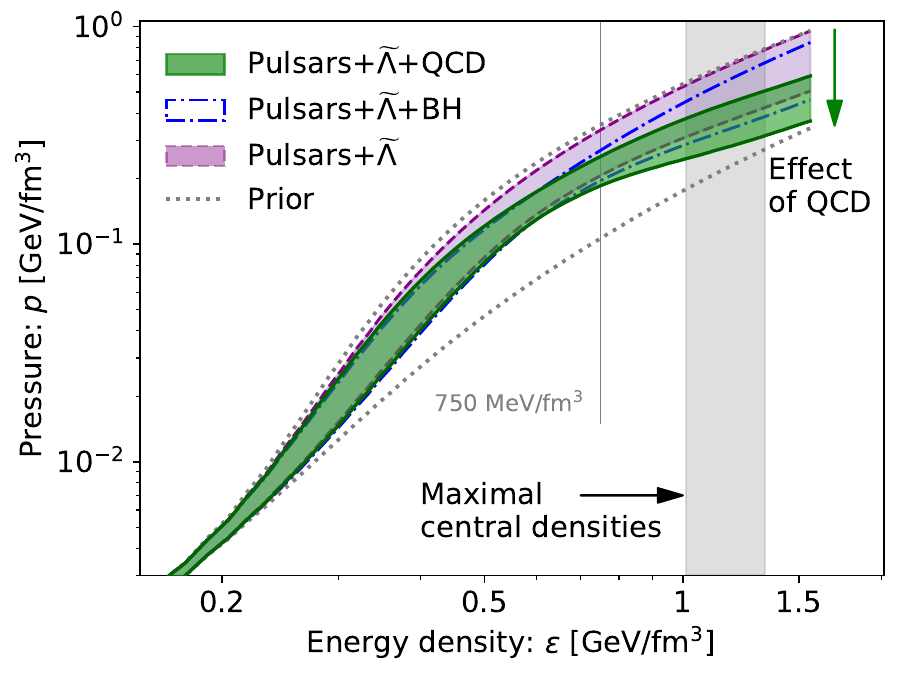}
    \caption{The impact of the QCD input on the EoS. The bands correspond to 67\%-credible intervals conditioned with different inputs. \emph{Pulsars+}$\widetilde\Lambda$ is conditioned with mass measurements of PSR J0348$+$0432 and PSR J1614$-$2230, with the combined mass and radius measurement of PSR J0740$+$6620, as well as the measurement of binary tidal deformability $\widetilde\Lambda$ from GW170817. \emph{Pulsars+$\widetilde\Lambda$+BH} further assumes that the binary merger product of GW170817 was a black hole. \emph{Pulsars+$\widetilde\Lambda$+QCD} additionally includes the QCD input, which excludes the largest pressures at high densities. The effect of the black-hole hypothesis is negligible if the QCD input is included. The grey band denotes a 67\%-credible interval for the maximal densities reached in stable non-rotating NSs. }
    \label{fig:e-p}
\end{figure}

The past years have seen great interest in the properties of ultradense nuclear matter in the cores of Neutron Stars (NSs). Driven by the fast progress in astronomical observations, many works have attempted to infer the properties of the equation of state (EoS) of NS matter using different combinations of observational and theoretical inputs. From the observational side, the EoS is most constrained by the discovery of massive NSs from pulsar timing \citep{Demorest:2010bx,Antoniadis:2013pzd,Cromartie:2019kug,Fonseca:2016tux,Fonseca:2021wxt}, combined mass and radius measurements from X-ray pulse profiling \citep{Riley:2019yda, Miller:2019cac,Riley:2021pdl,Miller:2021qha}, as well as measurements of the tidal deformability from gravitational-wave observations of binary NS mergers \citep{LIGOScientific:2017vwq, LIGOScientific:2018cki,LIGOScientific:2018hze,LIGOScientific:2020aai} and their associated electromagnetic counterparts \citep{LIGOScientific:2017zic}.

In addition to inputs from NS observations, the EoS is also informed by \emph{ab-initio} theoretical calculations. On the one hand, calculations within the framework of chiral effective theory (CET) constrain the EoS at low baryon densities $n$, i.e.,\ at and around the nuclear saturation density $n \lesssim 1.1 n_s$, where $n_s \approx 0.16$~fm$^{-3}$ \citep{Hebeler:2013nza,Drischler:2020yad}. On the other hand, the EoS can also be computed directly from the fundamental theory of Quantum Chromodynamics (QCD). These computations become reliable at large densities $n \gtrsim 40 n_s$, well above the densities reached in NSs \citep{Freedman:1976ub,Fraga:2004gz,Kurkela:2009gj,Kurkela:2016was,Gorda:2018gpy,Gorda:2021znl,Gorda:2021kme,Gorda:2021gha}.

The inference of the EoS from these astrophysical observations and theoretical inputs often proceeds by forming large ensembles of model-agnostic EoSs that are constrained or conditioned using the different inputs \citep{Hebeler:2013nza,Kurkela:2014vha}.
The majority of these EoS-inference setups have used different parametric \citep{Hebeler:2013nza, Tews:2018iwm,Dietrich:2020efo,Capano:2019eae,Raaijmakers:2019dks,Al-Mamun:2020vzu,Miller:2019nzo,Huth:2021bsp,Lim:2022fap} or non-parametric extensions \citep{Landry:2018prl, Landry:2020vaw,Essick:2019ldf,Miller:2021qha,Essick:2021kjb} of the CET EoS to NS star densities $n \sim 5-10n_s$.
There are also fewer works that additionally include the fundamental, high-density QCD constraint and interpolate through the full density range from 1.1$n_s$ to $40n_s$ \citep{Most:2018hfd,Altiparmak:2022bke,Kurkela:2014vha,Annala:2017llu,Annala:2019puf,Annala:2021gom, Jiang:2022tps}.

The results from analysis with and without the QCD input differ.
Notably, the works with the QCD input report a strong softening of the EoS and lower speed of sounds at energy densities of the order ${\epsilon \sim 750}$~MeV/fm$^3$, not discernible in works without the QCD input nor present in nuclear models. These features have been interpreted as the onset of a quark-matter phase in \citet{Annala:2019puf}.
It is, however, an open question whether these features occurring at NS densities are genuine predictions of QCD or whether they arise from attempting to interpolate the EoS through two orders of magnitude in density with a too restrictive interpolation function. Whether the QCD computations offer a significant and robust constraint to the EoS at NS densities determines whether QCD should be included in any complete EoS-inference setup. 

Here, we propose a new strategy to incorporate the QCD input that avoids the pitfalls of the interpolation. 
This advancement is made possible by 
a recent work \citep{Komoltsev:2021jzg} that demonstrated that
the requirement of causality and stability of the EoS imposes global constraints that feed the high-density QCD information down to lower densities in a completely model-independent way. It was found that the EoS is impacted by QCD down to $n \sim 2.2 n_s$ before including NS observations. What is not known, however, is whether QCD offers significant information beyond the constraints already provided by NS observations.

In this work we construct a simple Bayesian-inference setup, which we use to study the interaction of the QCD input with the NS observations. We extrapolate the CET EoS to $n = 10 n_s$ using a Gaussian-process (GP) regression and condition the prior using NS observations with or without QCD input. Our results are exemplified in \cref{fig:e-p}, which demonstrates the additional information gained by including the QCD input in conditioning the pressure $p$ as a function of energy density $\epsilon$ especially at high densities, leading to the aforementioned softening seen in previous works.

\section{Setup}
In this Section, we describe our framework based on GP regression that forms our prior $P(\EoS)$. Governed by Bayes' theorem, the prior is conditioned with the data to form a posterior process
\begin{equation}
P(\EoS \,|\, {\rm data} ) = \frac{ P(\EoS) \, P(  {\rm data} \,|\, \EoS \,)}{P({\rm data})},
\end{equation}
where $P({\rm data} \,|\,  \EoS)$ is the likelihood of an EoS given the data. 

Here we use data that is provided by astrophysical observations as well as from theoretical calculations in QCD.
These inputs are independent of each other and the likelihood factorises into 4 terms:
\begin{align}
P({\rm data} & \,|\, \EoS ) =  P({\rm QCD} \,|\, \EoS)  P( {\rm Mass} \,|\, \EoS)  \nonumber \\
 \times & P({\rm NICER} \,|\, \EoS) P({\rm \tilde \Lambda}, {\rm BH}  \,|\, \EoS).
 \label{eq:likelihoods}
\end{align}
These likelihoods correspond to the aforementioned QCD input, mass measurements of PSR J0348$+$0432 and PSR J1624$-$2230 using pulsar timing, the simultaneous X-ray measurement of mass and radius of PSR J0740$+$6620 using the NICER telescope, the tidal deformability measurement of GW170817 by LIGO/Virgo as well as the hypothesis supported by the electromagnetic counterpart of GW170817 that the resulting binary merger product is a black hole (BH). In many of the figures we consider the effect of pulsar measurements together and we denote 
\begin{equation}
P({\rm Pulsars} \,|\, {\rm EoS }) =  P( {\rm Mass} \,|\, \EoS)  P({\rm NICER} \,|\, \EoS)
\end{equation}

We now discuss the prior and the novel  QCD-likelihood function in detail, and give a brief description of how we implement each of the astrophysical measurements. 

\subsection{\texorpdfstring{Prior: $P(\EoS) $ }{Prior: P(EoS)}}

\begin{figure}
    \centering
    \includegraphics[width = 0.45 \textwidth]{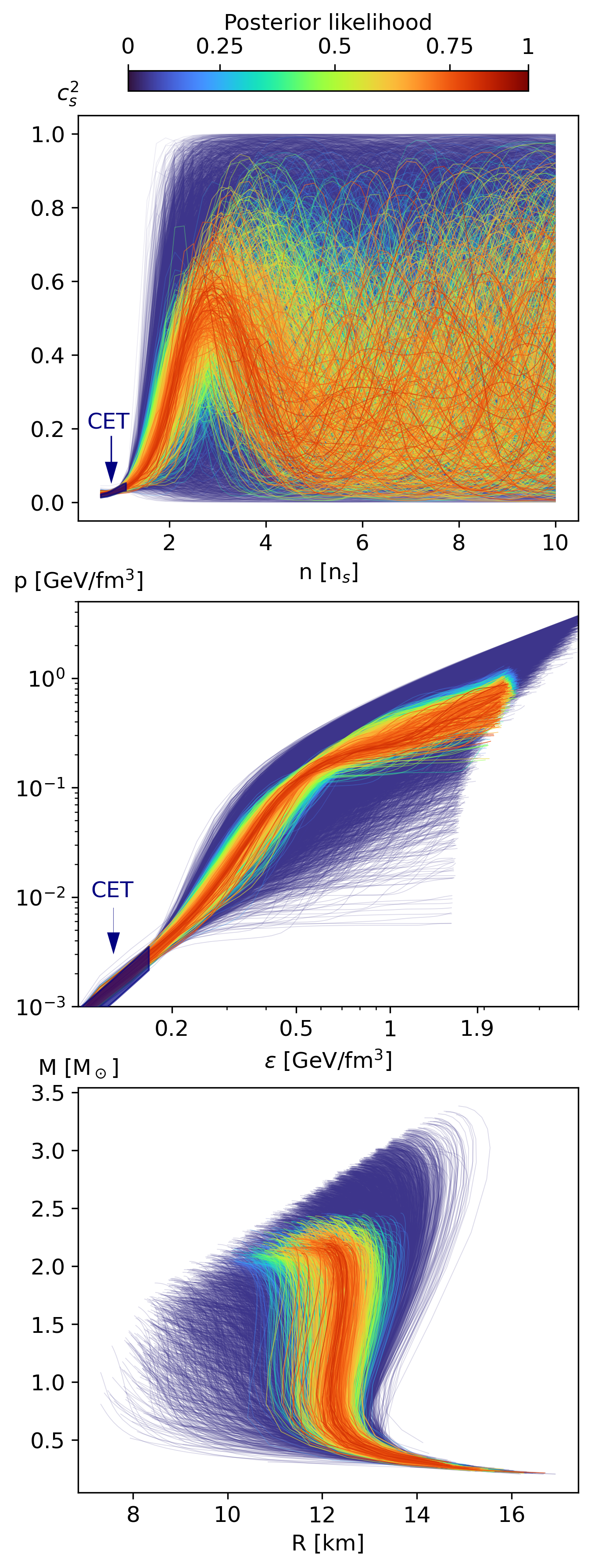}
    \caption{A sample of 10k EoSs drawn from the prior distribution conditioned with the low-density CET EoS. The coloring of the individual EoSs corresponds to the likelihood assigned according to \cref{eq:likelihoods}, imposing the information arising from all NS observations and the QCD input (i.e., Pulsars+$\widetilde\Lambda$+BH+QCD).}
    \label{cs2prior}
\end{figure}
In this section we describe our prior, which consists of an ensemble of EoSs that extrapolate the low-density EoS to relevant NS densities, up to ${n_L \equiv 10n_s}$. We choose to extrapolate the EoS using a GP regression \citep{Landry:2018prl,Essick:2019ldf} which is conditioned with an ($\beta$-equilibrated) EoS computed in CET up to $n = 1.1n_s$ \citep{Hebeler:2013nza}. For densities below $n = 0.57 n_s$ we merge the GP to the BPS crust EoS \citep{Baym:1971pw}.

As discussed in \citet{Komoltsev:2021jzg}, in order to make full use of the \emph{ab-initio} calculations in QCD, it is necessary to reconstruct the full form of the EoS, pressure as a function of number density $p(n)$, instead of using only its reduced form as a function of energy density $p(\epsilon)$. For this reason, our GP regression differs from that of \citet{Landry:2018prl,Essick:2019ldf}. Namely, we choose to extrapolate an auxiliary variable related to the speed of sound $c_s$,
\begin{equation}
    \phi(n) = -\log\bigl( 1/c_s^{2}(n) - 1\bigr)
\end{equation}
as a function of baryon number density $n$ instead of as function of energy density $\epsilon$. The logarithm in the above expression maps the causal range of $c_s^2 \in [0,1]$ to the range $\phi(n) \in [-\infty,\infty]$, suitable for a GP. 

Before conditioning the GP, we choose $\phi(n)$ to be normally distributed 
\begin{equation}
    \phi(n)  \sim \mathcal{N}\left( -\log \bigl( 1/\bar{c}_s^{2} - 1 \bigr), K(n,n') \right),
\end{equation}
with a Gaussian kernel $K(n,n') = \eta e^{-(n-n')^2/2 l^2}$. The three hyperparameters---the variance $\eta $, the correlation length $l$, and the mean speed of sound squared $\bar c_s^2$---determine the shape of the EoS where no data are available. To allow for a large variety of EoSs in our prior, we construct a hierarchical model by drawing the hyperparameters themselves from judiciously chosen random distributions
\begin{align}
l &\sim \mathcal{N}\bigl( 1.0 n_s, (0.25 n_s)^2 \bigr), \quad 
\eta \sim \mathcal{N}(1.25, 0.2^2), \nonumber \\
\bar c_s^2 &\sim \mathcal{N}(0.5, 0.25^2).
\end{align}

The GP is then conditioned with the CET EoS, leading to the $c_s^2(n)$ displayed in \cref{cs2prior}. We have taken the average and the difference of the ``soft" and ``stiff" EoS from \citet{Hebeler:2013nza} as the mean and 90\%-credible interval of the training data used to condition the EoS.

After conditioning, we draw a sample of EoSs from the GP and solve for the speed of sound $c_s$, the baryon chemical potential $\mu$, the energy density $\epsilon$
\begin{align}
    c_s^2(n) & = \frac{1}{e^{-\phi(n)}+ 1} \\
    \mu(n) & = \mu_0 \exp\left(\int_{n_0}^{n} dn' c_s^2(n')/n' \right) \\
    \epsilon( n) & = \epsilon_0 + \int_{n_0}^n d n' \mu(n'),
\end{align}
and the pressure $p(n) = - \epsilon(n) + \mu(n) n$. 

We then construct sequences of non-rotating stars ($M(n_{\rm central}), R(n_{\rm central}), \Lambda(n_{\rm central}) $) by solving the (perturbed) Tolman-Oppenheimer-Volkoff  (TOV) equations \citep{Hinderer:2007mb,Postnikov:2010yn,Han:2018mtj} for mass $M$, radius $R$, and tidal deformability $\Lambda$, with each EoS and for central densities up  $n_{\rm central} =  10 n_s$. We identify the stable branches and find the maximal masses $M_{\rm TOV}$ supported by each EoS.

\subsection{\texorpdfstring{QCD information: $P$(QCD$\,|\,$EoS)}{QCD information: P(QCD | EoS)}}
\label{sec:qcd_likelihood}
\begin{figure*}[ht!]
    \centering
    \includegraphics[width=1.\textwidth,trim={3cm 9cm 2cm 10cm},clip]{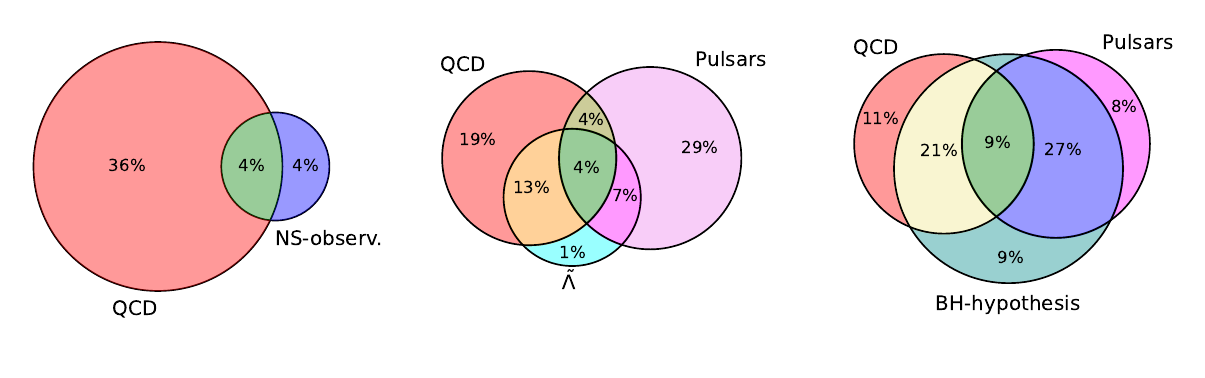}
    \caption{Overlap of various EoS constraints. The different areas refer to different posterior processes conditioned with the stated NS or QCD input, which provide a number of EoSs consistent with the stated input. The stated percentages correspond to the fraction of the prior EoSs in each posterior sample (resampled from the prior with frequencies proportional to likelihood weights, see main text for details). \emph{NS-observ.} corresponds to the all combined NS observations discussed here, including the mass and radius measurements as well as the tidal deformability measurement and BH hypothesis of GW170817.
    \label{fig:venn} 
    }
\end{figure*}

At high densities the asymptotic freedom of QCD allows one to access the EoS directly using perturbative calculations in QCD. These calculations give access to the full thermodynamic grand canonical potential as a function of baryon number chemical potential $\Omega(\mu)$, from which one can extract all other thermodynamic quantities. For our purposes, the most important physical quantities are the pressure $p$ and the number density $n$ evaluated at a given baryon chemical potential $\mu$, which we group into a vector ${\vec \beta} \equiv \{ p, n, \mu\}$. 
The current state-of-the-art result for the QCD grand canonical potential is the partial next-to-next-to-next-to-leading order (N3LO) perturbative-QCD computation of \citep{Gorda:2021kme, Gorda:2021znl}. 

As is the case with any perturbative calculation, it is important to estimate the missing-higher-order (MHO) error arising from the truncation of the perturbative series at a finite order. 
This truncation renders the result dependent on a residual, unphysical renormalization scale $\bar{\Lambda}$, which should be taken to be proportional to the physical scale in the problem to avoid large logarithms in the result. In the case of QCD at high density, one should take $\bar{\Lambda} \propto \mu$, but the optimal coefficient is undetermined. Formally, the dependence on the renormalization scale is of the order of the first uncalculated term in series.
This observation motivates the standard practice of estimating the missing-higher-order (MHO) terms by varying the renormalization scale around a fiducial scale by some fixed factor, though a Bayesian interpretation of the errors have been recently discussed in e.g.\  \citet{Cacciari:2011ze, Duhr:2021mfd}\footnote{The references \citep{Cacciari:2011ze, Duhr:2021mfd} also discuss in detail a machine-learning-based technique to estimate the convergence of the higher-order coefficients in the series. In our case the MHO errors are dominated by the scale variation and not the convergence, justifying our considering only the scale-variation error.}.

Here, we adopt the scale-averaging interpretation of \citet{Duhr:2021mfd} and interpret the perturbative result as a family of independent predictions for the correlated set ${{\vec \beta}_\mathrm{QCD}(X) = \{ p_\mathrm{QCD}(\mu_H, X), n_\mathrm{QCD}(\mu_H,X), \mu_H \}}$. This correlated set is evaluated at a fixed scale $\mu_H$, parameterized by the (dimensionless) renormalization scale $X \equiv 3\bar \Lambda/(2\mu_H)$.  The joint probability distribution for the set $\beta$ at the scale $\mu_H$, defined as $\beta_H$, is then obtained through a weighted average
\begin{align}
    P ({\vec \beta}_H) = \int d(\log X) w(\log X) \delta^{(3)}({ \vec \beta_H }- {\vec \beta}_{\rm QCD}(X)).
\label{eq:prior_beta_H}
\end{align}

In  \citet{Cacciari:2011ze} it was argued that a physically motivated choice for the weight function $w$ is a log-uniform distribution around a central scale
\begin{align}
    w(\log X) =  {\mathbf 1}_{[ \log(1/2),\,\log(2)]}(\log X),
\end{align}
with $\mathbf{1}_S$ the indicator function on a set $S$. We adopt this distribution in the following. The range $X \in [1/2,2]$---while in principle arbitrary---is suggested by phenomenological models \citep{Schneider:2003uz,Rebhan:2003wn,Cassing:2007nb,Gardim:2009mt} as well as the limit of QCD with a large number of flavors \citep{Ipp:2003jy}. Moreover, this range has been extensively used in previous studies and contains the minimally sensitive scales (defined through $\partial_X f(X)=0$, for $f(X)$) of $p_\mathrm{QCD}(\mu_H,X)$, $n_\mathrm{QCD}(\mu_H, X)$ and $c^2_{s,\mathrm{QCD}}(\mu_H, X)$ of the partial N3LO result used here.  

As for the value of $\mu_H$, we wish to use this perturbative information at a density that is as low as possible while staying in the regime where the MHO errors are under control. From these considerations, we adopt the scale $\mu_H = 2.6$~GeV, which was chosen in \citet{Fraga:2013qra}, because the uncertainty estimation at this value is similar to that of CET at 1.1n$_s$ (roughly $\pm24\%$ variation around mean value). This has also been the standard choice in subsequent perturbative QCD calculations \citep{Gorda:2018gpy,Gorda:2021znl,Gorda:2021kme}.

We now discuss how these QCD results can be used to construct a likelihood function that utilizes the perturbative calculation in a region where it is reliable but that can be imposed on an EoS at lower densities. The requirement that the triplet $\vec\beta$ be able to reach the QCD values at high densities using a stable and causal EoS imposes non-trivial constraints on the allowed values these quantities can take at NS densities. Here, we implement these constraints by excluding any GP extrapolations whose endpoint $\vec \beta_L = \{ p_L, n_L, \mu_L\}$\footnote{Note that here the $\vec\beta$-triplet with subscript $L$ indicates the last provided point of the EoS, while in \citet{Komoltsev:2021jzg} the subscript $L$ denoted the CET calculation. The reason is that we are effectively taking the last provided point of the EoS as the new low-density point.}, with $n_L \equiv 10n_s,$ cannot be connected to the corresponding QCD values with a stable and causal EoS. This requirement ensures that the EoS is consistent with the high-density limit also at all lower densities (assuming of course that the EoS is causal and stable up to $\vec\beta_L$). This very conservative requirement in effect allows the EoS to behave in a completely arbitrary way between the end-point of the GP extrapolation and the density region where perturbative-QCD results are assumed to be reliable. In particular, it is allowed to have an arbitrary number of phase transition with arbitrary latent heats and may have arbitrarily large segments where the speed of sound is near the speed of light (without exceeding it). 

\begin{figure*}[ht!]
    \centering
    \includegraphics[width=1.\textwidth]{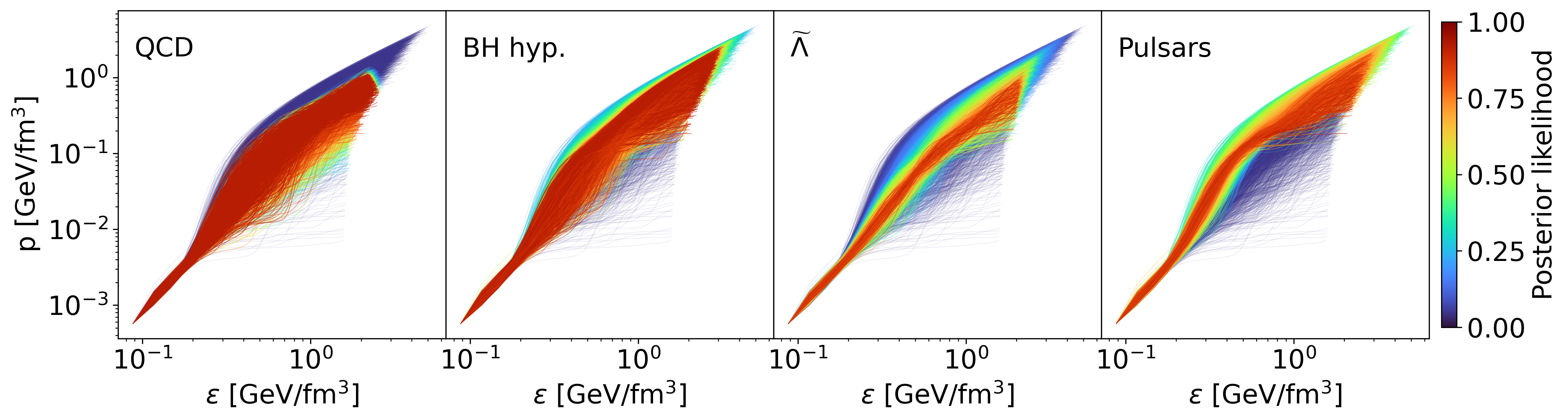}
    \caption{Impact of various inputs to the posterior $\epsilon$-$p$ process on a sample of 5k EoSs. The coloring of the individual EoSs corresponds to the likelihood assigned according to \cref{eq:likelihoods}, imposing the information arising from the different constraints.}
    \label{fig:ep-separate}
\end{figure*}

As discussed in \citet{Komoltsev:2021jzg}, these constraints can be implemented in a robust way without specifying the form of the EoS in the range $n_L < n < n_H$. This is done by comparing the pressure difference between the last point of the extrapolated EoS and the perturbative-QCD result $\Delta p = p_H - p_L$ to the minimal and maximal pressure differences $\Delta p_{\rm min}$ and $\Delta p_{\rm max}$ that can be reached with a thermodynamically stable EoS whose speed of sound squared does not exceed $ c_{\rm s, lim}^2$. For a given $\vec\beta_L$ the minimal and maximal pressure differences are \citep{Komoltsev:2021jzg}
{
\allowdisplaybreaks
\begin{align}
    \Delta p_{\rm min}(c_\mathrm{s,lim}^2) & = \frac{c_{\rm s, lim}^2}{1 + c_{\rm s, lim}^2 }
    \left( \mu_H \left(\frac{\mu_H}{\mu_L}\right)^{1/c_{\rm s, lim}^2} - \mu_L \right) n_L,\\
     \Delta p_{\rm max}(c_\mathrm{s,lim}^2) & = \frac{c_{\rm s, lim}^2}{1 + c_{\rm s, lim}^2 }
    \left( \mu_H  - \mu_L \left(\frac{\mu_L}{\mu_H} \right)^{1/c_{\rm s, lim}^2} \right) n_H.
\end{align}
}%
Note that the functions $\Delta p_\mathrm{min}$ and $\Delta p_\mathrm{max}$ implicitly depend on $\mu_L$ and $\mu_H$, though we do not show this dependence in their arguments. Furthermore, when taking $c_\mathrm{s,lim}^2 = 1$, we shall drop the argument for brevity.
The most conservative QCD likelihood function is then constructed by assign a zero Bayesian weight to any EoS whose $\vec\beta_L$ leads to the condition
$\Delta p \notin [ \Delta p_{\rm min},  \Delta p_{\rm max}]$. 

Accounting for the scale-variation error, the QCD likelihood function reads 
\begin{align}
P({\rm QCD} \,|\, {\rm EoS} ) = \int d \vec{\beta}_H P(\vec \beta_H) {\mathbf 1}_{[\Delta p_{\rm min}, \Delta p_{\rm max}]}(\Delta p).
\label{eq:QCDprior}
\end{align}
This integral is evaluated by substituting in \cref{eq:prior_beta_H} and integrating over $\vec \beta_H$ to resolve the $\delta$-function. In practice we perform 
Monte-Carlo integration by randomly drawing $X$ values from the distribution $w(\log X)$ and counting the frequency of the last extrapolated point $\vec\beta_L$ satisfying the condition for $\Delta p$.

The above conservative likelihood function does not in any way favor physically motivated EoSs in the density range $n_L <  n < n_H$ but merely asks if any causal interpolation exists. We may add further assumptions to the likelihood function by giving a higher weight to EoSs that can be connected to the QCD regime with an interpolation function that does not necessarily need to be extreme. One possible way to do so is to favor EoSs which can be connected to the QCD regime with EoSs that do not need to exceed some given limiting value of the speed of sound squared $c^2_{\rm s, lim}  < 1$ 
\begin{align}
     P({\rm QCD} \,|\, {\rm EoS} ) & = \int d c_{\rm s, lim}^2 P_0(c_{\rm s, lim}^2)
     \int d \vec\beta_H P(\vec \beta_H) \nonumber \\ 
    & \times {\mathbf 1}_{[\Delta p_{\rm min}(c_{\rm s,lim}^2),\Delta p_{\rm max}( c_{\rm s,lim}^2)]}(\Delta p),
\end{align}
where $P_0(c_{\rm s, lim}^2)$ is a prior distribution. 
 Given that the QCD value for $c_s^2$ is very close to its conformal value $c_s^2 = 1/3$, we may average the weight over different values of $c_{\rm s, lim}^2$
with, e.g., a flat prior function for the maximal speed of sound reached between $n_L< n < n_H$
\begin{align}
    P_0(c_{\rm s,lim}^2) = {\mathbf 1}_{[1/3,1]}(c_{\rm s,lim}^2).
\end{align}

In this work we will concentrate solely on the most conservative prediction given by \cref{eq:QCDprior} with $c_{\rm s,lim}^2 = 1$ and leave the exploration of the less-conservative likelihood functions for future studies.

\subsection{Astrophysical observations}

\begin{figure*}[ht!]
    \centering
    \includegraphics[width = 0.45\textwidth ]{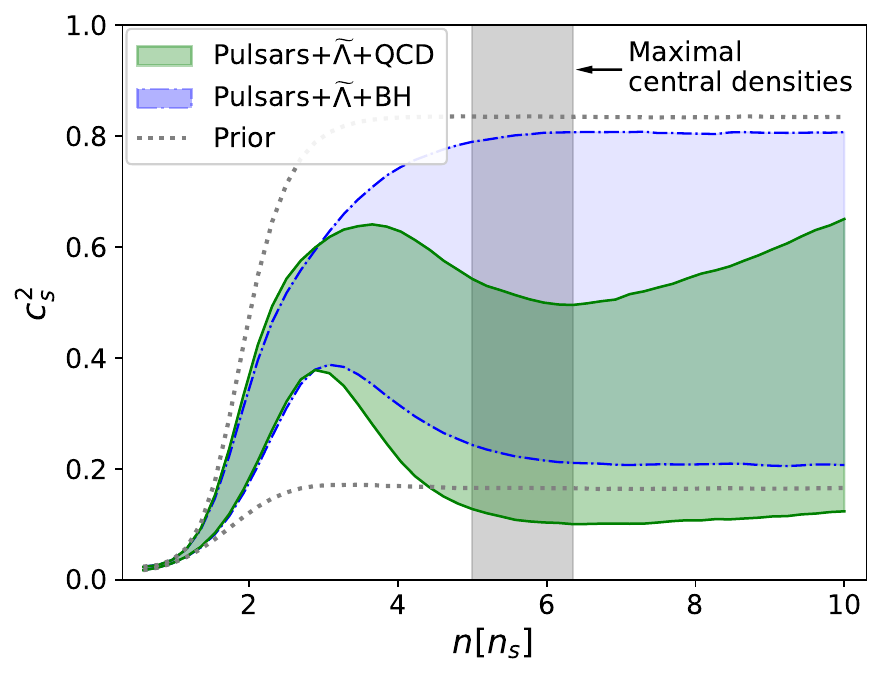}
    \includegraphics[width = 0.415\textwidth ]{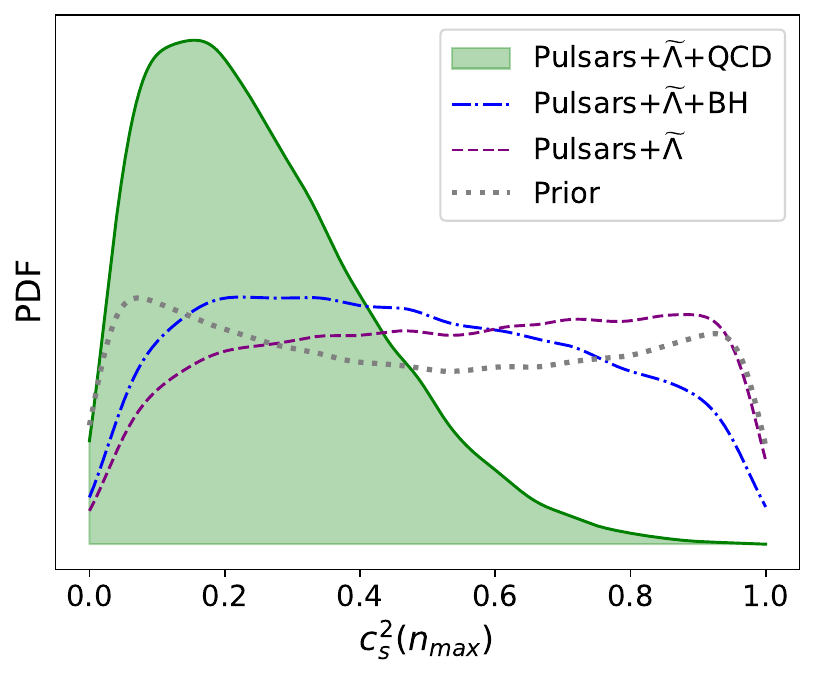}
    \caption{Effect of conditioning the prior of speed of sound with either the astrophysical observations alone or additionally with QCD. The bands correspond to 67\%-credible intervals. The NS observations require the EoS to be stiff at low densities, up to around 3$n_s$ while the QCD input favors smaller sound speeds at higher densities.  }
    \label{fig:cs2}
\end{figure*}

Here we briefly describe our implementations of the likelihoods for various NS measurements. We denote the various classes of observations ${\bf d}$, and construct our likelihood functions by marginalizing over additional source-specific model parameters that we must introduce to characterize the properties of the stars. These in principle depend on the EoS and stellar populations, as well as on selection effects. In particular, in the following we will assume a flat prior distribution for the masses of stars, $M_i$, up to the EoS-dependent maximal mass $M_{\rm TOV}$ for each star $i$ considered 
\begin{equation} 
P_0(M_i \,|\, {\rm EoS} ) = \frac{\mathbf{1}_{[M_{\rm min}, M_{\rm TOV}]}(M_i)}{M_{\rm TOV} - M_{\rm min} },
\label{eq:mass-prior}
\end{equation}
with $M_{\rm min} = 0.5 M_\odot$. With this, our full prior is given by $P_0(M_i \,|\, {\rm EoS} ) P({\rm EoS})$.
For a clear and more complete discussion of this point, we refer the reader to, e.g., \citet{Miller:2019nzo,Landry:2020vaw}%
\footnote{We note that the two references differ in the treatment of the normalization term of the prior mass distribution; \citet{Miller:2019nzo} omits the $M_{\rm TOV}$-dependent normalization in \cref{eq:mass-prior}. We have tested both priors and find a small effect on the posterior distribution. We show all the results following \citet{Landry:2020vaw} because it turns out that the effect from the prior is qualitatively similar to the QCD input. Hence, for studying the effect of QCD, taking the normalization of \citet{Landry:2020vaw} is a more conservative choice.}.

\subsubsection{ Mass measurements }
We require that the maximal mass supported by the EoSs, $M_{\rm TOV}$, is larger than the largest observed NS masses. We consider PSR J0348$+$0432 with  $M = 2.01\pm0.04 M_\odot$ \citep{Antoniadis:2013pzd} and PSR J1614$-$2230 with $M = 1.928\pm 0.017 M_\odot$ (in order to avoid double counting with the NICER measurements we do not include PSR J0740$+$6620).  For each of these stars $i$, we approximate the (per-source nuisance-marginalized) likelihood functions for the mass of the star $M_i$, $P({\bf d_{M}}\,|\, M_i)$, given the measurements ${\bf d_{M}}$ as a normal distribution with central value $M_i$ and standard deviation $\sigma_i$. We then give zero weight to an EoS if it cannot support a star of mass $M_i$, giving 
\begin{align}
P&({\rm Mass}_i  \,|\, {\rm EoS} )  \propto 
\int  dM_i P(d_{\bf M} \,|\, M_i) P_0(M_i \,|\, {\rm EoS}) \nonumber  \\
&\approx \frac{1}{2 (M_{\rm TOV} - M_{\rm min})} \left(
1
+ {\rm Erf}\left(\frac{M_{\rm TOV} - M_i}{\sqrt{2  }\sigma_i}\right)\right).
\end{align}
The final likelihood is then the product of two likelihoods for PSR J0348$+$0432 and PSR J1614$-$2230.

\begin{figure*}[ht]
    \centering
    \includegraphics[width=1.03\textwidth]{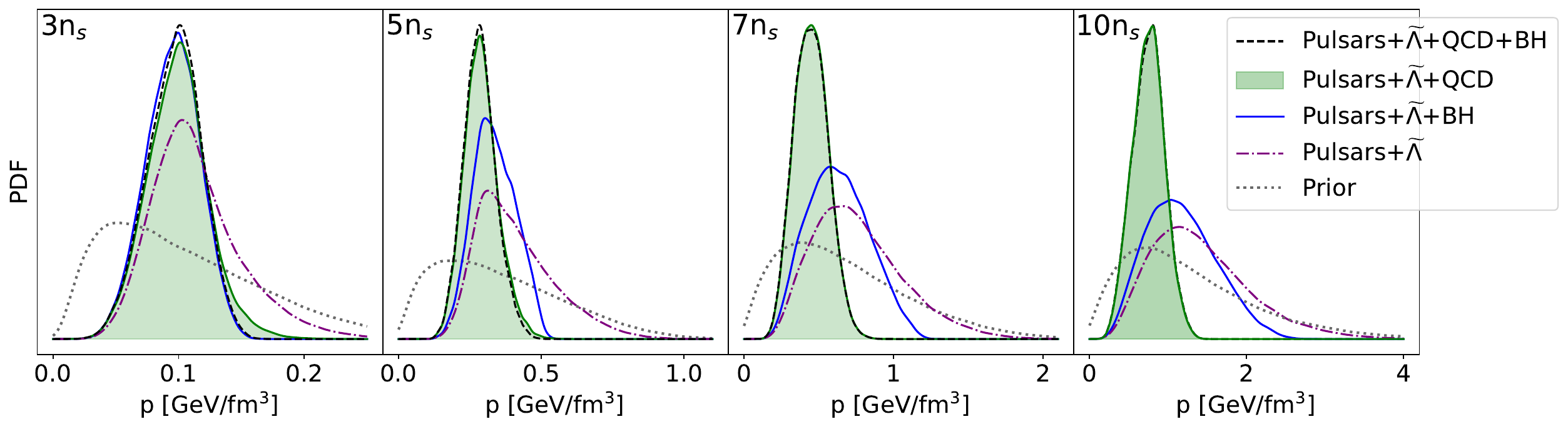}
    \caption{Kernel density estimate of the distribution of pressure at different densities. The filled green areas correspond to the posterior including all the NS observations and the QCD input (Pulsars+$\tilde \Lambda $+QCD+BH). The BH hypothesis adds negligible information after imposing the QCD input (Pulsars+$\tilde \Lambda$+QCD). }
    \label{fig:ep-kde}
\end{figure*}
\subsubsection{ Radius measurements }

In the case of PSR J0740$+$6620, we construct our likelihood function using the (nuisance-marginalized) joint posterior distribution for mass and (equatorial circumferential) radius $P( {\bf d_{\rm NICER}} \,|\, M,R)$ given the observations ${\bf d_{\rm NICER}}$ determined using combined measurements of  NICER and XMM-Newton observatories (Fig.~1 (right) of \citet{Miller:2021qha})
\begin{align}
P(&{\rm NICER}\,|\, {\rm EoS} ) \propto \nonumber \\  &\int dM_i P( {\bf d_{NICER}} \,|\, M_i,R(M_i) ) P_0(M_i \,|\, {\rm EoS}). 
\end{align}
\subsubsection{ Tidal deformability }

The LIGO/Virgo Collaborations give the accurate measurement of the chirp mass 
\begin{equation}
    \mathcal{M}_\mathrm{chirp} \equiv \frac{(M_1 M_2)^{3/5}}{(M_1 + M_2)^{1/5}} = 1.186(1) M_\odot.
\end{equation}
as well as the joint posterior density function $P({\bf d_{GW}}  \,|\, \widetilde \Lambda, q)$ (see Fig.\ 12 of \citet{LIGOScientific:2018hze}, low-spin prior) for the binary-tidal-deformability parameter $\widetilde \Lambda$ and mass ratio of the binary merger components $q = M_1/M_2$, marginalized over all the other nuisance parameters (including the chirp mass). 

For this measurement, we require a joint mass prior for the binary $P_0(M_1, M_2 \,|\, \EoS)$, which we take as
\begin{equation}
    P_0(M_1, M_2 \,|\,\EoS) = \frac{\mathbf{1}_{[M_2, M_\mathrm{TOV}]}(M_1) \mathbf{1}_{[M_\mathrm{min}, M_\mathrm{TOV}]}(M_2)}{1/2(M_\mathrm{TOV} - M_\mathrm{min})^2},
\end{equation}
which is uniform on the set of possible binaries with $M_1, M_2 \in [M_\mathrm{min}, M_\mathrm{TOV}]$ and $M_1 \geq M_2$. Given the accurate measurement of $\mathcal{M}_{\rm chirp}$, choosing a value for $M_1$ effectively enforces a delta-function on $M_2$, determining the mass of the second merger component $M_2 = M_2( M_1, \mathcal{M}_{\rm chirp})$.  We then construct the likelihood function using the expression%
\begin{align}
P(\widetilde \Lambda \,|\, {\rm EoS} ) =  &\int_{M_1 > M_2} \!\!\!\!\!\! dM_1 
P_0(M_1, M_2 \,|\, \EoS) \nonumber \\
&\times P( {\bf d_{\rm GW}} \,|\, \widetilde \Lambda , q(M_1, M_2)).
\label{eq:TD} 
\end{align}
Here, the integral may be calculated by integrating from the $q = 1$ value of $M_2 \approx 1.362(1) M_\odot$.

\subsubsection{ Black-Hole hypothesis }\label{sec:BH-hyp}

The final piece of observational input that we implement in our analysis is the hypothesis that the remnant in GW170817 collapsed to a BH.
This conclusion is based on the short gamma-ray burst GRB170817A \citep{LIGOScientific:2017ync} associated with GW170817, whose properties are, in light of current numerical modelling of NS mergers, consistent with the formation of a BH~\citep{Margalit:2017dij,Rezzolla:2017aly,Ruiz:2017due,Shibata:2017xdx,Shibata:2019ctb}. To implement this hypothesis, we proceed as follows.

Consider a possible initial binary with a mass ratio $q$ for a fixed EoS. Throughout the merger process, the total baryon number of the system $N(q)$ is conserved. Hence, $N_1 + N_2 = N_\mathrm{remnant}(q) + N_\mathrm{ejecta}(q)$, where $N_1$ and $N_2$ are the baryon numbers of the two binary components.  In order for the remnant to collapse to a BH for this value of $q$, it must be case that $N_\mathrm{remnant}(q) > N_\mathrm{TOV}$, the baryon number of the maximum stable (non-rotating) star. Following \citet{Annala:2021gom}, we ignore the small mass ejecta and consider a more conservative assumption that $N(q) > N_\mathrm{TOV}$.  To implement this condition within our Bayesian framework, we take as the likelihood (cf.~\cref{eq:TD})
\begin{align}
P&({\rm BH} \,|\, {\rm EoS} ) = \int_{M_1 > M_2} \!\!\!\!\!\! dM_1 P_0( M_1, M_2 | {\rm EoS}) \nonumber\\
&\times P({\bf d_{GW}}  \,|\, q(M_1, M_2)  ) \mathbf{1}_{[N_\mathrm{TOV}, \infty]}\bigl( N(q) \bigr) , 
\label{eq:BH_alone}
\end{align}
where $ P({\bf d_{GW}}  \,|\, q  )$ is the 1d marginal posterior distribution for low-spin priors from \citet{LIGOScientific:2018hze}. As above, here $q$ is fixed by $M_1$ and $\mathcal{M}_\mathrm{chirp}$.

This condition is however not independent from the constraint from the tidal deformability and both likelihoods depend
on the same mass ratio $q$ since it concerns the same event. Hence when using the BH hypothesis, we must impose both inputs simultaneously by defining the likelihood as
\begin{align}
P&(\widetilde \Lambda, {\rm BH} \,|\, {\rm EoS} ) = \int_{M_1 > M_2} dM_1 P_0( M_1, M_2 \,|\, {\rm EoS}) \nonumber \\
&\times P( {\bf d_{\rm GW}} \,|\, \widetilde \Lambda , q(M_1, M_2) )\mathbf{1}_{[N_\mathrm{TOV}, \infty]}\bigl( N(q) \bigr).
\label{eq:BH_with_TD}
\end{align}

\begin{figure*}
    \centering
    \includegraphics[width = 0.5 \textwidth]{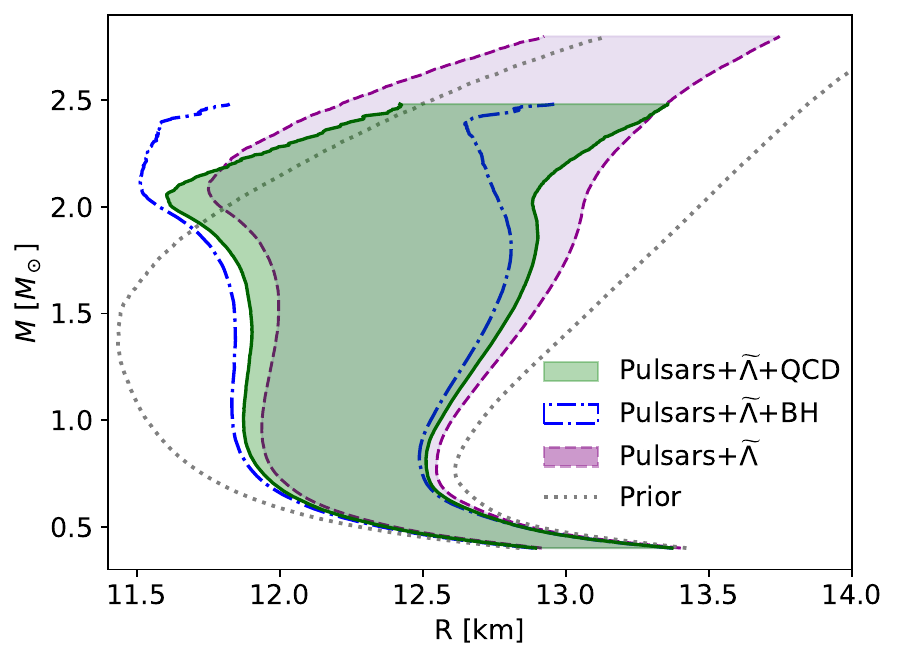}
    \includegraphics[width = 0.45 \textwidth]{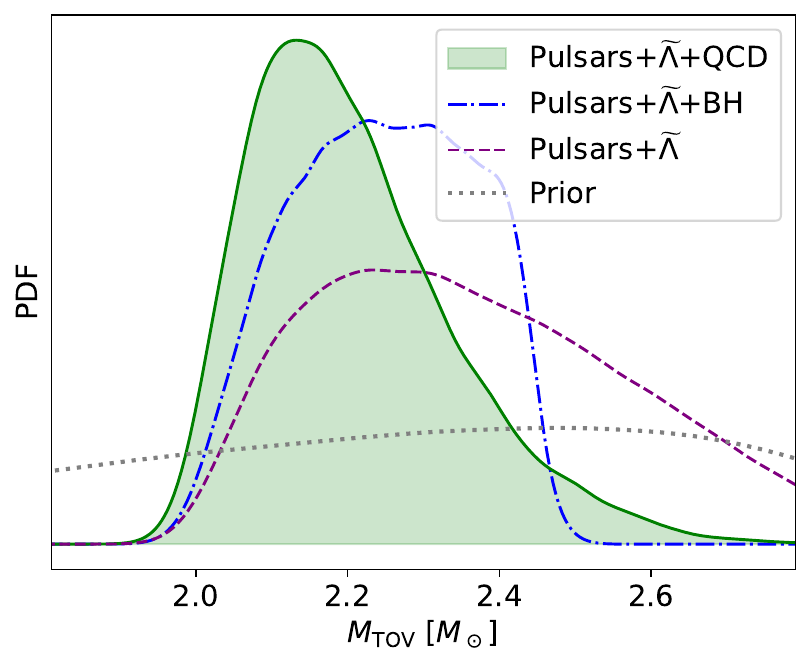}
    \caption{(Left) The 67\%-posterior density contours of the radius $R$ for different $M$. (Right) Marginalized posterior density of the maximal mass of a stable non-rotating NS star. The main effect of the QCD input in the $M$-$R$ plane is to reduce the maximal masses reached in stable NSs. Note that in the left figure we have cut off the contours at high $M$ where there are too few EoSs to form a smooth boundary.}
    \label{fig:MR}
\end{figure*}

\section{Results}
We have generated a prior sample of 120k EoSs\footnote{We performed the initial analysis with an ensemble size of 20k EoSs and checked results for stability by including an additional 100k EoSs. No qualitative difference is observed.} from the prior process, and for each of these EoS we have computed the corresponding likelihood functions appearing in \cref{eq:likelihoods}. Before turning to physical observables, we first discuss the extent to which the different constraints from the observations and QCD input either complement or corroborate each other. We do so by generating multiple posterior distributions, each conditioned with a subset of the inputs, and compare the overlap of these distributions.

In practice, for each selected subset of inputs, we resample the prior ensemble with an accept/reject step with probability proportional to the  likelihood of the EoS, 
\begin{equation}
\frac{P({\rm data} \,|\, {\rm EoS} )}{\max_{\rm EoS}[P({\rm data} \,|\, {\rm EoS} )]},
\end{equation} 
where here the likelihood includes only the selected inputs (these resampled ensembles are used only in \cref{fig:venn}). The fraction of EoSs in such a posterior sample is then proportional to the marginal probability $P({\rm data})$ and illustrates how constraining the selected inputs are. We then examine the resulting posterior samples for different sets of selected inputs: If two posterior samples contain a large number of common EoSs, then the two observables corroborate each other, and the two inputs give similar information about the EoS; if the intersection is small, the two inputs offer complementary information about the EoS. Below, we report the fraction of EoSs remaining in a given posterior, normalized by the number of prior EoSs. An alternative way of characterizing this information would be e.g., using the Kullback-Leibler divergence as done in  \citet{Riley:2019yda,Raaijmakers:2019qny}. However, we find the comparison of posterior samples more intuitive.

From our prior, 8\% of the EoSs are consistent with combined NS observations (including the BH formation hypothesis). Similarly, 40\% of the prior sample are consistent with QCD (see \cref{fig:venn}a). Most importantly, only 4\% of the EoSs are simultaneously allowed by both the NS observations and QCD input; that is, imposing the QCD input significantly constrains the EoS by removing half of the otherwise allowed EoSs in the posterior sample. 
 
Further analyzing the overlap between different measurements, we find unsurprisingly that the pulsar mass measurements and the NICER measurement of PSR J0740+6620 give strongly overlapping information, with the latter being more constraining due to the information on the radius (not shown). We also find that the LIGO measurement of $\tilde \Lambda$, the combined pulsar measurements, and the QCD input all offer complementary information to one another, illustrated in \cref{fig:venn}b. Finally, we find that the input that most corroborates the QCD input is the BH hypothesis, see \cref{fig:venn}c. 
In fact, we see that after imposing the pulsar observations, all EoSs consistent with QCD are also consistent with the BH hypothesis in the resampled posterior distribution, indicating that the formation of a BH in GW170817 is a pre/postdiction of QCD. 
The effect of the BH hypothesis is negligible if the QCD input is included. 

Moving on to physical observables, we show the impact of various constraints on the $\epsilon$-$p$ process in \cref{fig:ep-separate}, which displays samples drawn from our prior ensemble and colored by the different likelihood functions. We note in passing that considering only the QCD input (panel a) uses no astrophysical inputs and is thus dependent on neither general relativity nor the assumption that NSs are composed entirely of Standard-Model matter. It therefore can be used as a starting point in studies testing these assumptions without using circular logic \citep{LopeOter:2019pcq, Lope-Oter:2021vxl}.

We see that the effect of the QCD input primarily decreases $p(\epsilon)$ for large $\epsilon$, softening the EoS, which is qualitatively different from the effect from the pulsar input (panel d), which favors EoSs that are sufficiently stiff and limits the pressure from below. 
The $\tilde\Lambda$ measurement from GW170817 also drives the $\epsilon$-$p$ process towards softer values but limits the pressure at significantly smaller densities than the QCD input. This illustrates the complementary nature of these inputs seen earlier in \cref{fig:venn}b. As expected from the above discussion, the BH hypothesis corroborates the QCD input in the qualitative conclusion that the EoS must not be too stiff at high densities.

The combined effect of the different inputs is shown in \cref{fig:e-p} above, where the prior is conditioned sequentially with different inputs. The QCD input offers a significant reduction of the uncertainty of the EoS, complementary to the astrophysical observations. This is particularly prominent if neglecting the BH hypothesis,
which relies on models of jet formation from BH+disc and NS+disc systems.
The effect of the QCD constraint becomes dominant at densities above $\epsilon \sim 750$~MeV/fm$^3$ ($450$~MeV/fm$^3$, if not including the BH hypothesis). Our result clearly demonstrates that the origin of the softening seen by previous studies utilizing the QCD constraint directly is indeed a robust prediction of QCD.

The gray band in \cref{fig:e-p} corresponds to the density range reached in the cores of $M_{\rm TOV}$ stars. From the figure it is clear that the QCD input affects densities that can be reached within NSs. This is even more clearly visible in \cref{fig:ep-kde} displaying a kernel density estimate of the pressure at four fixed energy densities. At the density $n = 3 n_s$ the QCD input mildly constrains the EoS, but starting already from $n = 5 n_s$ the effect of the QCD input is striking. 

The softening due to QCD is also reflected in the posterior speed of sound, displayed in \cref{fig:cs2}. The NS observations primarily exclude EoSs that are too soft in the density range $n \sim 2-4 n_s$. At higher densities, the observations offer little information, and the posterior closely follows the prior. Imposing the QCD input however dramatically reduces the variation at densities $n \gtrsim 3 n_s$ forcing the equation of state to have speed of sound $c_s^2 \lesssim 0.6$. This is perhaps most dramatically seen in the distribution of the speed of sound in the centers of $M_{\rm TOV}$ stars, shown in \cref{fig:cs2}. Here, the QCD input strongly disfavors large speeds of sounds at the highest densities reached in NSs. Note that no information about the speed of sound of high-density QCD, $c_s^2 \lesssim 1/3$, is used in the conditioning. 

Despite these changes on the $\epsilon$-$p$ and $c_s^2$ processes, the QCD input has only a small effect in the $M$-$R$ plane, as seen in \cref{fig:MR}a. This renders it difficult to impose similar constraints on the EoS using $M$-$R$ -measurements only. The main effect of the QCD input in the $M$-$R$ plane is to reduce the maximal masses reached in stable NSs, see \cref{fig:MR}b.

\begin{figure}
    \centering
    \includegraphics[width=0.48\textwidth]{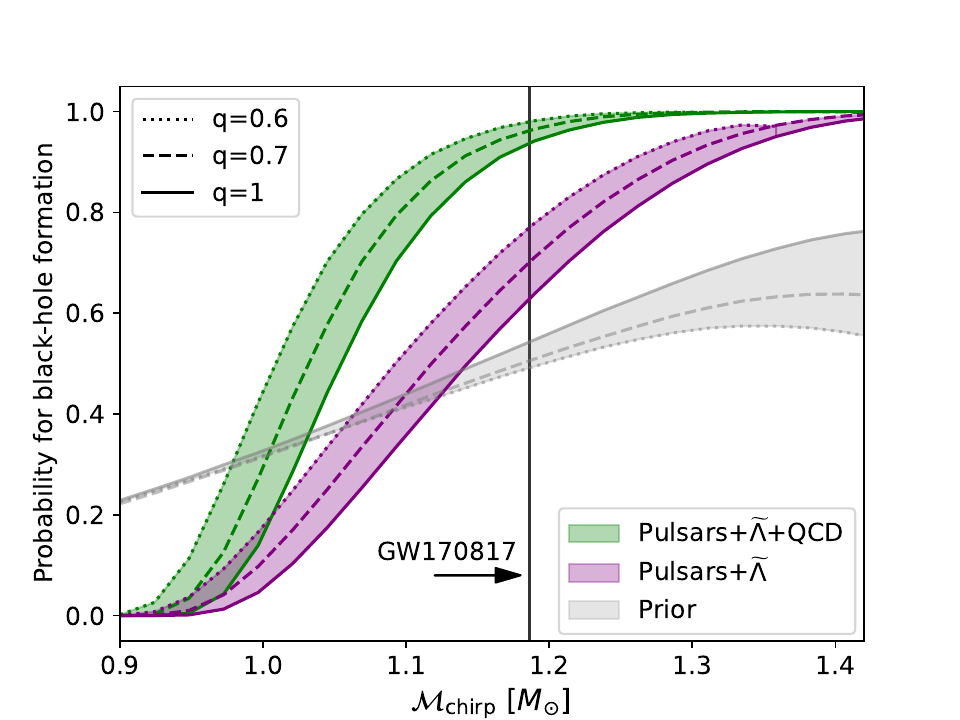}
    \caption{The posterior probability that a binary-NS merger with chirp mass $\mathcal{M}_{\rm chirp}$ and mass ratio $q$ leads to a BH. The purple area shows the prediction using information from pulsar and gravitational-wave observations but does not make an assumption about the binary merger product of GW170817. Including the QCD input (green area) leads to the prediction that a binary merger with $\mathcal{M}_{\rm chirp} = 1.186$ leads to a  BH with 90\% credence. }
    \label{fig:bh_pred}
\end{figure}

As remarked above, after imposing the pulsar and $\tilde \Lambda$ inputs, the QCD input implies that the binary merger product in GW170817 was a BH. Similar predictions can be made for other potential binary-NS events with different mass configurations. We conduct the same analysis as in Sec.~\ref{sec:BH-hyp} above for different potential binaries characterized by parameters $\mathcal{M}_\mathrm{chirp}$ and $q$, and show our results in \cref{fig:bh_pred}. We find that the QCD input enhances the posterior probability of forming a BH in any given event, e.g.~predicting that binary collisions of equal-mass NSs will produce a BH with greater than $95\%$ ($68\%$) credence for masses $M \geq 1.38 M_\odot$ ($M \geq 1.25 M_\odot$).

Finally, we note that while our prior ensemble contains a large number of EoSs that admit twin stars with multiple stable branches, the mass measurements give a vanishing weight to EoSs with multiple stable branches. We note however, that this finding is not in contradiction with other works [e.g.~\citet{Benic:2014jia,Christian:2019qer,Christian:2021uhd}] finding twin-star solutions. These works utilize more specific EoS parametrizations including explicit strong first-order transitions, which are approximated by smooth but rapidly changing functions and have a small prior weight in our analysis.

\section{Conclusions}
In this work, we have characterized the impact from \emph{ab-initio} calculations of the part of the Standard Model of particle physics describing the strong nuclear force, Quantum Chromodynamics, on the inference of the equation of state of ultradense matter in neutron-star cores. We have demonstrated that the QCD input significantly constrains the equation of state even after all the current astrophysical observations are taken into account, reducing uncertainties in the determination of the equation of state particularly at the highest densities reached in neutron stars. We found that imposing the QCD constraint requires the equation of state to significantly soften at energy densities $\epsilon \gtrsim 450$~MeV/fm$^3$. 

Our work offers an explanation for the discrepancy between works that have observed the softening and those which have not as a result of either including the QCD input or not. By using the most conservative possible construction to impose the QCD input, we have demonstrated that this softening is a robust prediction of the fundamental theory of the strong interactions.

Of the observational inputs we have discussed, the black-hole formation  hypothesis has most overlap with QCD, with QCD giving a much stronger constraint of the two. 
The hypothesis requires non-trivial astrophysical modelling of the formation of the jet whose reliability may be difficult to quantify. Therefore is it is particularly fortunate that the QCD input offers an independent constraint supporting the conclusion arising from the black-hole hypothesis. Similarly, future multimessenger observations of binary mergers with different masses offers a possible way to empirically test the QCD prediction. 

We emphasize the importance of including all possible inputs when inferring the equation of state. All the measurements and theoretical calculations come with different, non-overlapping systematics and uncertainties. Therefore, a convincing global understanding of the matter in the most extreme conditions is only reached through combining multiple, mutually consistent inputs. 

Lastly, we note that a strategy that encompasses all the possible inputs is the one most likely to find a conflict between them. Assuming that all the systematics are under control, such a discrepancy would signal a failure of the underlying assumptions that NSs are described by the Standard Model and general relativity. This underlines the importance of including all possible reliable inputs in order fully exploit these extreme objects as a tool for fundamental discovery.

In order to facilitate an easy implementation of the QCD input to other, possibly more sophisticated, inference setups, we provide the QCD likelihood function used in this work in an easy-to-use Python implementation located on Zenodo \citep{komoltsev_oleg_2023_7781233}\footnote{The QCD likelihood function is also available on GitHub} \citep{OlegGithub}. 

Finally, after completing this study, another work appeared studying the impact of the QCD input on the NS EoS inference \citep{Somasundaram:2022ztm}. We are able to reproduce their results within our analysis by applying the constraints from \citet{Komoltsev:2021jzg} at the maximum central density reached along a given EoS, rather than $10 n_s$, and they are able to reproduce our results within their analysis when applying the constraints at $10 n_s$ \citep{somasundaram_priv_com}. %
Pushing the construction defined in Sec.~\ref{sec:qcd_likelihood} down to lower density obviously decreases the constraining power of the QCD input. We note that while this can be done, it requires the EoS to have significantly different behavior between the maximal central densities reached in NSs and $10n_s$, compared to the prior used in this work and others. In particular, it requires a very strong phase-transition-like behavior and significant density ranges where the speed of sound is close to the speed of light. While this is interesting in its own right, we note that the effect of such phase-transition behavior is also not included for the low densities (for the first systematic studies see \citet{Gorda:2022lsk}).

\section*{Acknowledgements}
The authors thank Eemeli Annala, Bjorn Auestad, Aleksas Mazeliauskas, Joonas Hirvonen, Alex Nielsen, Joonas N\"attil\"a, Sanjay Reddy, Achim Schwenk, and Aleksi Vuorinen for useful discussions. The authors especially thank Tore Kleppe for discussions about Gaussian processes.
This work is supported in part by the Deutsche Forschungsgemeinschaft
(DFG, German Research Foundation) -- Project-ID 279384907 -- SFB 1245
and by the State of Hesse within the Research Cluster ELEMENTS 
(Project ID 500/10.006). The authors are listed in alphabetical order by last name.

\bibliography{main}

\end{document}